# Colossal Type-II Multiferroic Polarization Driven by Collinear Spin Orders

Chengxi Huang[1,⊥], Xinhai Tu[2,⊥], Jintao Jiang[1], Xiangang Wan[2,3,4,5*], Erjun Kan[1,*]

[1] *MIIT Key Laboratory of Semiconductor Microstructure and Quantum Sensing, Nanjing University of Science and Technology, Nanjing, Jiangsu 210037, P. R. China.*

[2] *National Laboratory of Solid State Microstructures and School of Physics, Nanjing University, Nanjing 210093, P. R. China.*

[3] *Collaborative Innovation Center of Advanced Microstructures, Nanjing University, Nanjing 210093, China*

[4] *Hefei National Laboratory, Hefei 230088,China*

[5] *Jiangsu Physical Science Research Center, Nanjing University, Nanjing 210093, China*

[⊥] These authors contributed equally to this work.

★Correspondence and requests for materials should be addressed to

E. K. (ekan@njust.edu.cn), X. W. (xgwan@nju.edu.cn)

# Abstract

Achieving strong magnetoelectric coupling (MEC) together with large ferroelectric polarization remains a central challenge in type-II multiferroics. In conventional spin-driven multiferroics, the induced polarization is usually mediated by spin-orbit coupling (SOC) or spin-lattice coupling (SLC). Since many representative systems are based on 3*d* transition-metal ions, where SOC is relatively weak and SLC-induced lattice distortions are often limited, their polarizations are typically much smaller than those of proper ferroelectrics. Moreover, electric polarizations in type-II multiferroics are generally induced by spiral spin orders stabilized by competing magnetic interactions, which often leads to relatively low magnetic transition temperatures. In this Letter, using spin-group symmetry, we propose an SOC- and SLC-independent route to MEC in collinear 3*d* magnetic systems. We show that, even for a noncentrosymmetric lattice structure, different collinear magnetic configurations can either forbid or allow electric polarization, indicating direct magnetic control of polarization and hence strong MEC. The first-principles calculations excluding SOC on monolayer 2H-$VS_2$ support this picture: a collinear stripy antiferromagnetic order induces an in-plane ferroelectric polarization up to 25.00 $\mu C/cm^2$, about two orders of magnitude larger than that of typical type-II multiferroics. Furthermore, our microscopic model suggests that the induced polarization originates from SOC-independent *p*-*d* hybridization governed by electronic hopping. Our results suggest a possible route toward type-II multiferroics combining strong MEC with large electronic polarization in collinear 3*d* magnetic systems.

*Introduction* Multiferroics and magnetoelectrics, in which magnetic and electric orders coexist and couple to each other, have attracted sustained interest because of both their fundamental significance and their potential applications in low-power spintronic and multifunctional devices [1-12]. In conventional type-I multiferroics [13-15], magnetism and ferroelectricity originate from distinct microscopic mechanisms, typically resulting in sizable electric polarization but weak magnetoelectric coupling (MEC). In contrast, type-II multiferroics exhibit intrinsically strong MEC because their polarization is directly induced by magnetic order itself [16-18]. However, a long-standing difficulty is that the polarization in type-II multiferroics is usually much smaller than that in type-I multiferroics [19-23].

The microscopic origin of this limitation can be traced to the mechanisms that generate spin-driven polarization. In conventional type-II multiferroics, the polarization is produced through mechanisms related to spin-orbit coupling (SOC), such as the spin-current [24] or inverse Dzyaloshinskii–Moriya (DM) mechanism [25] and SOC-mediated *p-d* hybridization [26,27]. These mechanisms are effective in coupling spin textures to electric polarization, but their magnitude is constrained by the relativistic energy scale of SOC. This limitation is particularly relevant for 3*d* transition-metal systems, where SOC is usually weak. Another important route is exchange striction [28-30], in which magnetic order induces lattice distortions through spin-lattice coupling (SLC). Although this mechanism does not require SOC, the resulting polarization is often restricted by relatively weak SLC and by the elastic energy cost of ionic displacement [31,32]. Therefore, conventional type-II multiferroics face a general bottleneck: they do exhibit strong MEC, but the induced polarization is often limited by either weak SOC or weak SLC. A further issue is that electric polarizations in many type-II multiferroics are induced by noncollinear spiral magnetic orders [6]. Such spin textures typically arise from competing exchange interactions, magnetic frustration, or relativistic anisotropic interactions [6,7]. While these ingredients are essential for generating noncollinear magnetism, they are commonly associated with relatively low magnetic ordering temperatures in many known type-II multiferroics [6-9]. This make it desirable to search for alternative mechanisms and materials in 3*d* magnetic systems where large ferroelectricity can be induced by simpler collinear magnetic orders without relying on either SOC or SLC.

In this Letter, motivated by the search for magnetic materials that may host both strong MEC and large ferroelectric polarization without relying on either SOC or SLC-induced ionic displacement, we employ spin-group (SG) symmetry to analyze possible magnetoelectric responses in two-dimensional (2D) collinear magnetic systems. Our symmetry analysis shows that different collinear magnetic configurations can either forbid or allow electric polarization even for a noncentrosymmetric lattice, indicating a strong coupling between spin order and polarization. Guided by this criterion, we identify typical 2H-$MX_2$ monolayers (M=3*d* transition metal; X=O, S, Se, Cl, Br, I) with noncentrosymmetric $D_{3h}$ point group as a promising platform. Taking 2H-$VS_2$ as a prototype, we find that the ferromagnetic (FM) state enforce a vanishing polarization. By contrast, the stripy antiferromagnetic (s-AFM) order introduces an in-plane electric polarization. Our first-principles calculations excluding SOC confirm this symmetry prediction and show that the polarization reaches 25.00 $\mu C/cm^2$, exceeding that of typical type-II multiferroics by two orders of magnitude and approaching the scale of proper ferroelectrics. Our microscopic model further reveals that this large spin-driven polarization indeed originates from SOC-independent *p-d* hybridization between ligand and transition-metal orbitals. In this mechanism, polarization is governed primarily by electronic hopping, and is therefore not constrained by the smallness of SOC. A substantial enhancement can be expected, because electronic hopping usually occurs on the eV scale and is generally much larger than SOC.

*Symmetry analysis* Traditionally, magnetic group has been employed to characterize the symmetry of magnetic systems. However, with the recent emergence of altermagnetism [33,34], SG, where spin and lattice degrees of freedom are not rigidly locked, has become a powerful framework for describing magnetic systems in the absence of SOC [35-40]. To elaborate how collinear magnetic configurations directly control electric polarization under the SG symmetry constraints, here we consider a simple noncentrosymmetric 2D triangular magnetic cluster as an example (Fig. 1). This triangular cluster belongs to the nonpolar $D_{3h}$ point group. As shown in Fig. 1a, the collinear FM configuration preserves SG operations $\{E||C_{3z}\}$ and $\{E||C_{2x}\}$, which together enforce a vanishing polarization despite the absence of inversion symmetry. However, for the ↑↑↓ magnetic state, the out-of-plane SG operation $\{E||C_{3z}\}$ is broken, allowing a finite polarization along the $x$ direction (Fig. 1b).

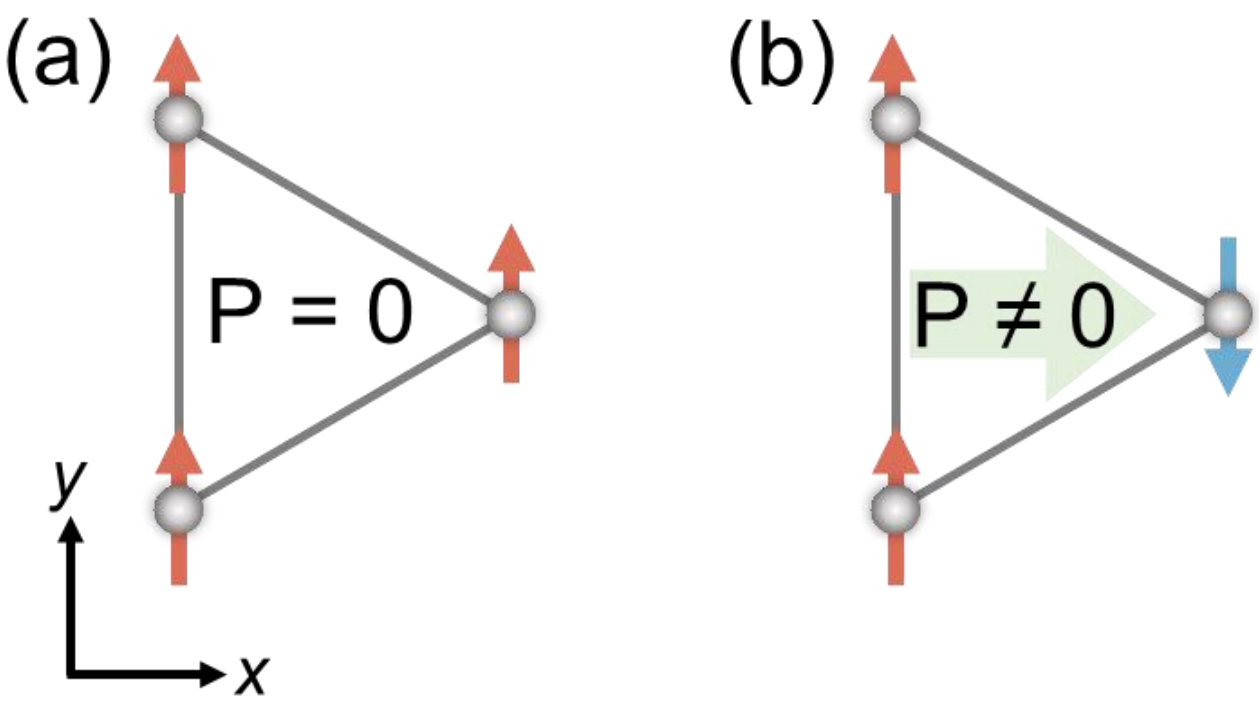

**Fig. 1.** Schematic diagrams of ferroelectricity driven by collinear spin order in triangle magnetic cluster without inversion symmetry. Red and blue arrows represent in-plane spin moments. Although the lattice itself lacks inversion symmetry, as discussed in the main text, the FM state remains nonpolar due to the relevant SG symmetry operations $\{E||C_{3z}\}$ and $\{E||C_{2x}\}$. By contrast, the collinear ↑↑↓ spin configuration breaks $\{E||C_{3z}\}$ and allowing a finite ferroelectric polarization.

*Material Calculations* One of the typical 2D material families that possess the noncentrosymmetric $D_{3h}$ point-group symmetry discussed above is the well-known 2H-phase transition-metal dichalcogenide monolayers with trigonal-prismatic coordination [41-43]. To exclude the SOC effect, we focus on the 3$d$ 2H-$MX_2$ monolayers (M=3$d$ transition metal; X=O, S, Se, Cl, Br, I), where the SOC strength is expected to be relatively weak. Our first-principles calculation results indicate that these materials can provide an ideal platform for realizing collinear type-II multiferroic candidates with strong MEC and large ferroelectric polarization without relying on SOC and SLC. Here, we take monolayer 2H-$VS_2$ as a representative example [44], where the magnetic V ions form a 2D triangular spin lattice (Fig. 2a). The details of calculation method are provided in section I of supplementary materials (SMs) [45]. In the absence of SOC (Unless otherwise stated, all calculations in the following are performed without SOC), the FM order in 2H-$VS_2$ preserves the $\{E||C_{3z}\}$ and $\{E||C_{2x}\}$ SG symmetry and is nonpolar. Hence, we define the spin-driven electric polarization ($\Delta P_\phi$) as the difference in polarization between a given spin order $\phi$ ($P_\phi$) and the nonpolar FM order ($P_{FM}$), namely, $\Delta P_\phi = P_\phi - P_{FM}$. Using the Berry phase method [51], our first-principles calculations show that the $P_{FM} = 0$ for 2H-$VS_2$ monolayer. In contrast, the s-AFM state possesses a nonzero polarization along the *x*-axis ($P_{s\text{-}AFM} \neq 0$), because it breaks the $\{E||C_{3z}\}$ SG symmetry, consistent with the spin group analysis presented in Fig. 1. Surprisingly, the magnitude of $\Delta P_{s\text{-}AFM}$ is remarkably large, reaching 25.00 μC/cm$^2$, which is about two orders of magnitude larger than that of common type-II multiferroics (e.g. ~0.08 μC/cm$^2$ in $TbMnO_3$ [16]) and is comparable to that of proper ferroelectrics (e.g. ~30 μC/cm$^2$ in $BaTiO_3$ [52]). We have also investigated the transition between the FM and s-AFM orders by gradually rotating one of the two spins in the rectangular unit cell from 0 to 180°. Figure 2b shows a continuous change of $\Delta P_\phi$ from 0 to 25.00 μC/cm$^2$ during this magnetic evolution, further demonstrating the spin-driven nature of the electric polarization. Including SOC leads to only a negligible change in $\Delta P_{s\text{-}AFM} \approx 24.98$ μC/cm$^2$, indicating that such a large spin-driven polarization is not governed by SOC. We further examine the possible role of SLC by fully relaxing the lattice in the s-AFM state, as shown in Fig. 2c. The resulting lattice distortion is found to be quite small, suggesting that the SLC effect is weak. In addition, this small structural distortion slightly reduces $\Delta P_{s\text{-}AFM}$ to 23.07 μC/cm$^2$, indicating that the large polarization is also not driven by SLC-induced ionic displacement.

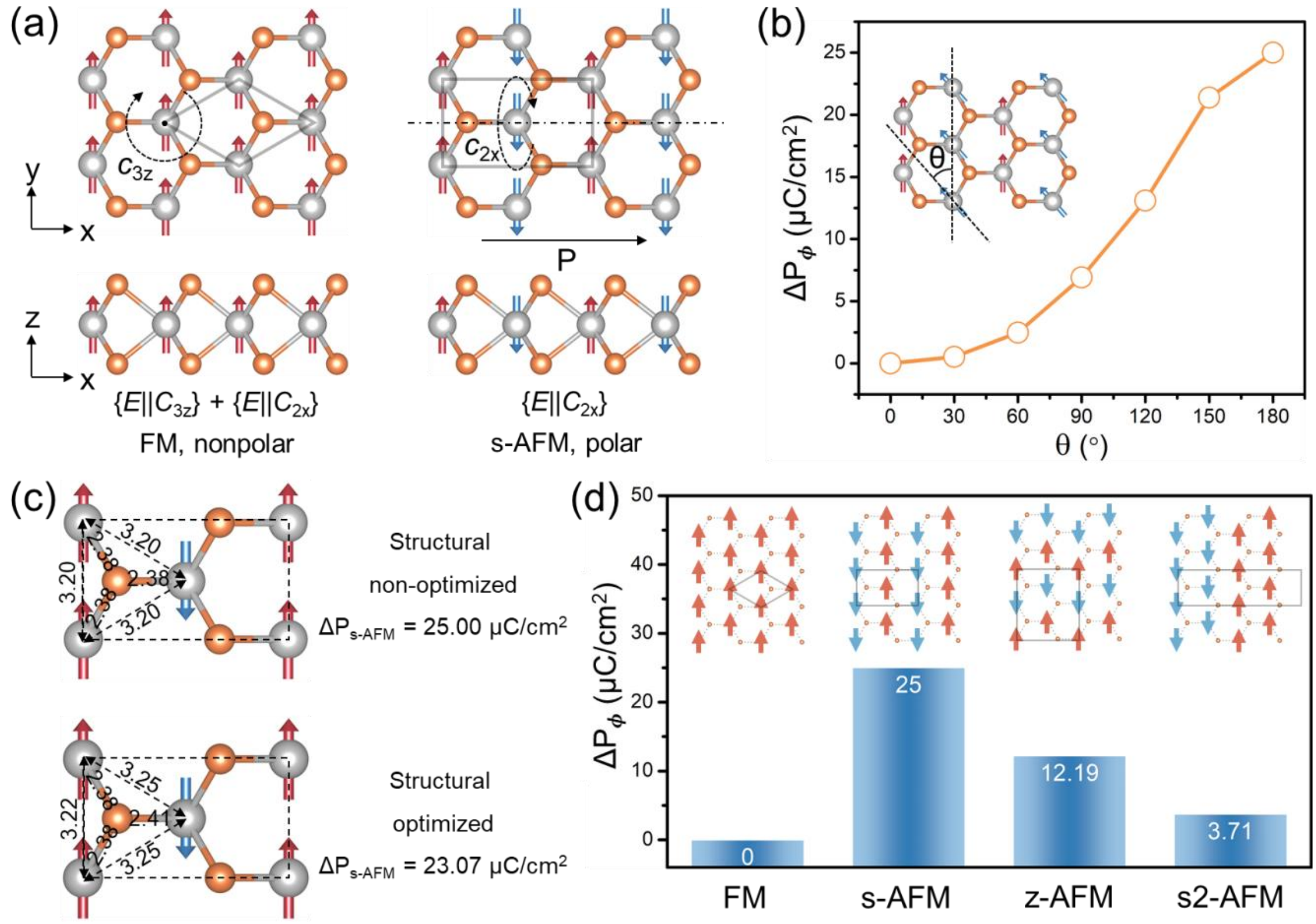

Fig. 2. (a) Top and side views of 2H-$VS_2$ monolayer in ferromagnetic (FM) and stripy antiferromagnetic (s-AFM) orders. Gray and orange balls represent V and S atoms, respectively. Red and blue arrows represent spins on V atoms. (b) Variation of the spin-driven electric polarization ($\Delta P_\phi = P_\phi - P_{FM}$) during the transition process from FM to s-AFM order by gradually rotating half of the spins without structural optimizations. (c) V-S and V-V bond lengths (in Å), and spin-driven electric polarization ($\Delta P_{s\text{-}AFM}$) for the s-AFM state without and with structural optimization. (d) $\Delta P_\phi$ for FM, s-AFM, zigzag antiferromagnetic (*z*-AFM) and ↑↑↓↓-stripy antiferromagnetic (s2-AFM) orders.

To avoid the ambiguity associated with the multivalued polarization in the Berry-phase method, we further verify the magnitude of $\Delta P_{s\text{-}AFM}$ using the finite electric field method [53,54] in section II of the SMs [45]. We have also investigated the spin-driven polarization in a 2D triangular system with open boundaries and the results show that the spin-driven electric polarization is robust against boundary conditions (see section III in SMs [45] for details). Besides the s-AFM order, other possible collinear spin orders can also exhibit spin-driven electric polarizations as long as they break the $\{E\|C_{3z}\}$ SG symmetry. Figure 2d shows the distinct magnitudes of $\Delta P_{\phi}$ for the s-AFM, zigzag AFM (z-AFM) and ↑↑↓↓-stripy AFM (s2-AFM) orders.

Similar magnetoelectric coupling behavior is also observed in other 2H-$MX_2$ monolayers (Table S1 in SMs [45]). The magnitude of $\Delta P_{s\text{-}AFM}$ for the 2H-$MX_2$ monolayers can reach up to ~94 μC/cm$^2$. Furthermore, an electric-field switch of magnetic order could be achieved due to the large $\Delta P_{\phi}$. For instance, although the 2H-$ScI_2$ is FM in its ground state, an in-plane electric field greater than 0.16 V/nm along the $x$-axis reverses the sign of $\Delta E(\varepsilon)$ [defined as the energy difference between s-AFM and FM states, namely $\Delta E(\varepsilon) = E_{s\text{-}AFM}(\varepsilon) - E_{FM}(\varepsilon)$] and induces a transition from FM to s-AFM state (Fig. S5 in SMs [45]). Such a coercive electric field ($\varepsilon_c$) required to switch the spin order between FM and s-AFM states can be estimated by the ratio between $\Delta E(\varepsilon=0)$ and $\Delta p_{s\text{-}AFM}$ (defined as the difference on electric dipole moment between s-AFM and FM states, namely $\Delta p_{s\text{-}AFM} = p_{s\text{-}AFM} - p_{FM}$), namely, $\varepsilon_c \approx \Delta E(\varepsilon=0)/\Delta p_{s\text{-}AFM}$. These results highlight the potential applications in spintronic devices for these materials.

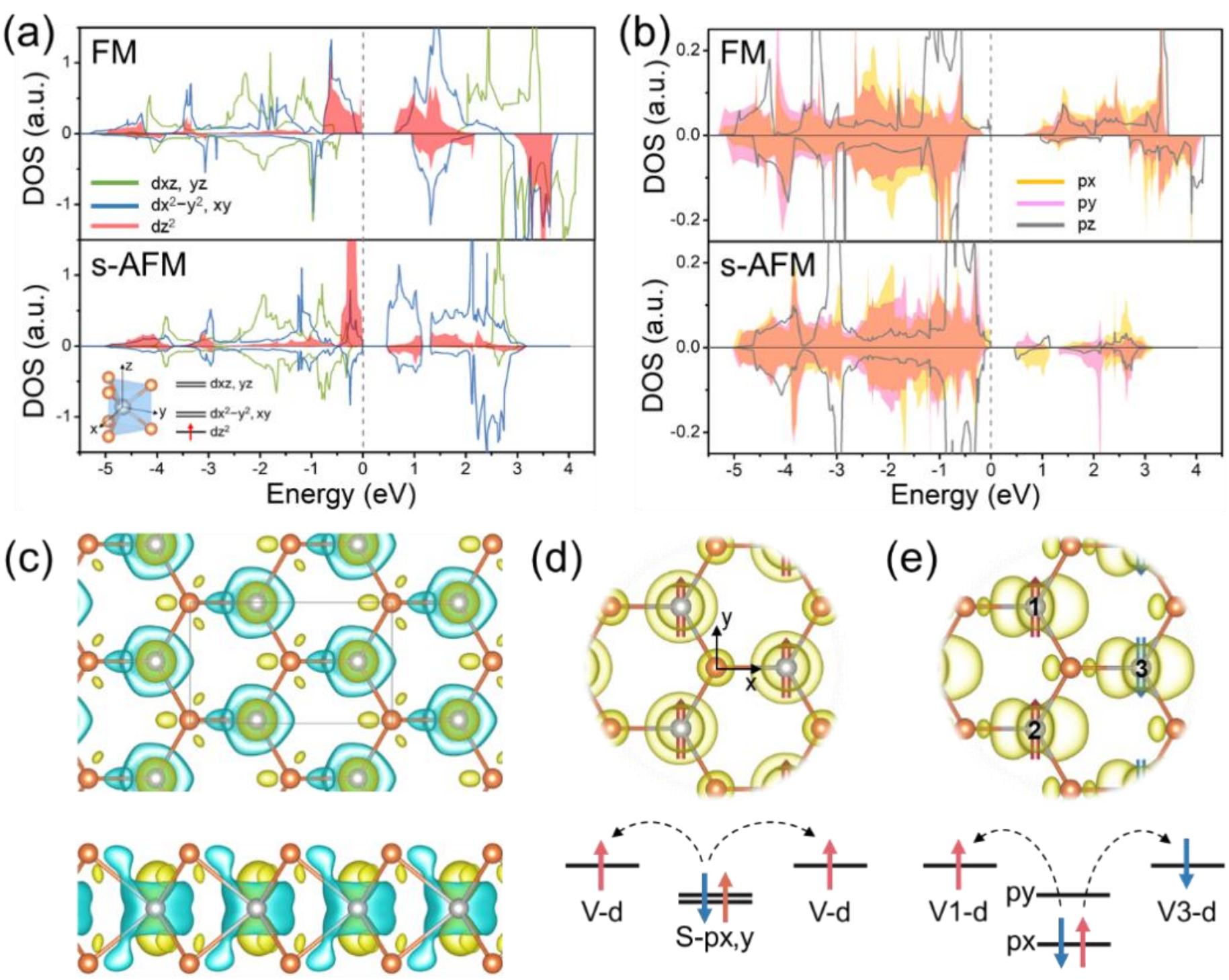

**Fig. 3.** Projected density of states (DOS) for (a) V-3$d$ and (b) S-3$p$ orbitals in ferromagnetic (FM) and stripy antiferromagnetic (s-AFM) states, respectively, for 2H-$VS_2$ monolayer. Inset: schematic diagram of the trigonal-prismatic crystal field and the resultant splitting of the V-3d orbitals. (c) Differential charge density between FM and s-AFM orders with identical atomic structure. Yellow and cyan isosurfaces represent charge accumulation and depletion. Partial charge density for the valence band maximum (upper panels) and schematics of superexchange interactions (lower panels) in (d) FM and (e) s-AFM orders. Red and blue arrows represent spins.

*Microscopic model* The SG symmetry determines whether the spin-driven ferroelectricity can emerge, but it cannot explain why the $\Delta P_{s\text{-}AFM}$ is such large. To elucidate the origin of this SOC-independent MEC, we construct a microscopic model for monolayer 2H-$VS_2$. In the $D_{3h}$ trigonal-prismatic crystal field formed by the surrounding six S ions, the V-3*d* orbitals split into one one-dimensional irreducible representation $\{d_{z^2}\}$, and two 2D irreducible representations $\{d_{x^2-y^2}, d_{xy}\}$ and $\{d_{xz}, d_{yz}\}$. The single *d* electron of each $V^{4+}$ ion occupies the lowest-energy $d_{z^2}$ orbital, giving rise to a formal magnetic moment of 1 $\mu_B$. The projected density of states in Fig. 3(a) shows that the electronic states near the Fermi level are dominated by the V-$d_{z^2}$ orbital, consistent with this crystal-field picture. Since the $d_{z^2}$ orbital carries magnetic quantum number $m$=0, SOC can be neglected to the leading order. The microscopic origin of the polarization can be understood from the spin-dependent *p-d* hybridization in a $V_3S$ triangular cluster. Figure 3(c) reveals a distinct charge disproportion around each S atom, which originates from the inequivalent superexchange processes in the ↑↑↓ spin triangle. In the ↑↑↑ state, the three V-S-V superexchange paths are equivalent and the $p_x$ and $p_y$ orbitals of S ion are nearly degenerate apart from small in-plane anisotropy [Fig. 3(d)]. However, in the ↑↑↓ or ↓↓↑ configuration, the V1-S-V2 superexchange channel mediated mainly by the S-$p_y$ orbital is suppressed in the parallel spin alignment, whereas the V1-S-V3 and V2-S-V3 superexchanges channels, mainly involving the S-$p_x$ orbital, are enhanced in the antiparallel spin alignment. This spin-dependent imbalance lifts the near degeneracy between $p_x$ and $p_y$ orbitals (Fig. 3e), leading to an anisotropic *p-d* orbital hybridization and a charge disproportion. As a result, an in-plane electronic dipole moment is generated. To quantify this mechanism, we construct the nearest-neighbor V-S hopping Hamiltonian $H_t$ as follow, constrained by the threefold rotation $C_{3z}$ and mirror symmetry $\sigma_y$ and $\sigma_z$:

$$
\begin{gathered}
H_t=\sum_\sigma \sqrt{3}tp_{1y\sigma}^\dagger d_{1z^2\sigma}-\sqrt{3}tp_{1y\sigma}^\dagger d_{2z^2\sigma}+tp_{1x\sigma}^\dagger d_{1z^2\sigma}+tp_{1x\sigma}^\dagger d_{2z^2\sigma}-2tp_{1x\sigma}d_{3z^2\sigma} \\
+t'p_{1z\sigma}^\dagger d_{1z^2\sigma}+t'p_{1z\sigma}^\dagger d_{2z^2\sigma}+t'p_{1z\sigma}^\dagger d_{3z^2\sigma} \\
+\sum_\sigma \sqrt{3}tp_{2y\sigma}^\dagger d_{1z^2\sigma}-\sqrt{3}tp_{2y\sigma}^\dagger d_{2z^2\sigma}+tp_{2x\sigma}^\dagger d_{1z^2\sigma}+tp_{2x\sigma}^\dagger d_{2z^2\sigma}-2tp_{2x\sigma}d_{3z^2\sigma} \\
-t'p_{2z\sigma}^\dagger d_{1z^2\sigma}-t'p_{2z\sigma}^\dagger d_{2z^2\sigma}-t'p_{2z\sigma}^\dagger d_{3z^2\sigma}+\mathrm{h}.c.
\end{gathered}
$$

where $\sigma$=↑↓ labels the spin and hopping amplitudes $t$ and $t'$ describe hybridization between V $d_{z^2}$ orbitals and S $p$ orbitals. The effective local exchange field acting on the V spins is described by

$$H_M=-J\sum_{a=1,2,3}m_a\cdot S_a, \tag{1}$$

where $J$ represents the Hund coupling energy [37]. Using the second-order degenerate perturbation theory, we obtain the perturbation ground state for the in-plane FM configuration [24]

$$|\psi_0^{FM}>=\frac{1}{\sqrt{3}}\sum_a|d_{a\downarrow}>-\frac{\sqrt{3}}{\Delta}t'(|p_{1z\downarrow}>-|p_{2z\downarrow}>), \tag{2}$$

which gives

$$\mathbf{P}=e<\psi_0^{FM}|\mathbf{r}|\psi_0^{FM}>=0. \tag{3}$$

This vanishing polarization is enforced by the spin-group operations $\{E||C_{3z}\}$ and $\{E||C_{2x}\}$, consistent with both our symmetry analysis and DFT results. In contrast, for the ↑↑↓ configuration, the ground-state wave function up to first order reads

$$
\begin{gathered}
|\psi_0^{AFM}>=|\psi_0^{AFM(0)}>+|\psi_0^{AFM(1)}> \\
=\frac{1}{\sqrt{2}}(|d_{1\downarrow}>+|d_{2\downarrow}>)-\frac{\sqrt{2}}{\Delta}[t(|p_{1x\downarrow}>+|p_{2x\downarrow}>)+t'(|p_{1z\downarrow}>-|p_{2z\downarrow}>)],
\end{gathered} \tag{4}
$$

leading to a finite polarization

$$\mathbf{P}=e<\psi_0^{AFM}|\mathbf{r}|\psi_0^{\mathrm{AFM}}>=-\frac{8e}{\Delta}(tI+t'I')x, \tag{5}$$

where $I$ and $I'$ are dipole matrix elements between V $d_{z^2}$ and S $p$ orbitals. Equation (5) shows that the polarization is controlled directly by the hopping amplitudes $t$ and $t'$, rather than by SOC or ionic displacements. Since hoppings are on the eV electronic energy scale, this mechanism naturally produces a much larger polarization than conventional SOC-mediated *p-d* hybridization mechanisms [26,27]. The integral matrix elements $I$ and $I'$ are approximately on the order of a few Bohr radii (i.e., a few Å) [24]. The $\Delta$, $t$ and $t'$ are estimated to be 3, 0.2 and 0.5 eV for monolayer 2H-$VS_2$, respectively, from the maximally localized Wannier functions analysis. These give an electric polarization reaching the order of 10 μC/cm$^2$, which is consistent with our DFT calculation results. Besides, the reversal of the magnetic configuration leaves the relative angle between spins at sites $i$ and $j$ invariant, thereby the direction of polarization remains unchanged. More details are shown in section VI of SMs.

*Conclusion*— In summary, we have proposed an SOC- and SLC-independent route to strong MEC and large ferroelectric polarization in collinear 3*d* magnetic systems. SG symmetry shows that different collinear magnetic configurations can either forbid or allow electric polarization even in a noncentrosymmetric lattice crystal, enabling direct magnetic control of polarization. Taking monolayer 2H-$VS_2$ as a prototype, our first-principles calculations confirm that the FM state is nonpolar, whereas the s-AFM state breaks the relevant SG symmetry and induces an in-plane polarization up to 25.00 μC/cm$^2$, about two orders of magnitude larger than typical type-II multiferroics. Our microscopic model attributes this large polarization to SOC-independent *p-d* hybridization governed by electronic hopping rather than SOC or SLC. These results suggest that collinear 3*d* magnets provide a promising platform for type-II multiferroicity combining strong MEC with large electronic polarization.

This work is supported by the Ministry of Science and Technology of the People´s Republic of China (No. 2025YFA1411001 and 2025YFA1411301), NSFC (T2125004, 12274227, U24A2010 and 12188101), Innovation Program for Quantum Science and Technology (Grant No. 2021ZD0301902), Quantum Science and Technology-National Science and Technology Major Project (Grant No. 2024ZD0300101), Natural Science Foundation of Jiangsu Province (Grants No. BK20233001, No. BK20243011), Fundamental and Interdisciplinary Disciplines Breakthrough Plan of the Ministry of Education of China (Grant No. JYB2025XDXM411), Fundamental Research Funds for the Central Universities (Grant No. KG202501), and the New Cornerstone Science Foundation. C.H. and E.K. acknowledge the support from the Tianjin supercomputer centre.

The data that support the findings of this article are openly available [55].